\documentclass[%
 reprint,
 amsmath,amssymb,
 aps,
 pra,
longbibliography,
]{revtex4-2}

\usepackage{booktabs}
\usepackage{multirow}
\usepackage{braket}
\usepackage{siunitx}[=v2]
\DeclareSIUnit\sample{S}
\usepackage{graphicx}% Include figure files
\usepackage{dcolumn}% Align table columns on decimal point
\usepackage{bm}% bold math
\usepackage{xcolor, soul}
\newcommand{\state}[3]{#1\text{#2}_{#3}}

\begin{document}

\preprint{APS/123-QED}

\title{Simultaneous Multi-Band Demodulation Using a Rydberg Atomic Sensor}% Force line breaks with \\
%Multi-tone Demodulation vs Multi-Band
%Multi-Band
%5-tone demodulation
%two-decatde
%condtinuous vs simultaneous
%
\author{David H. Meyer}
\email{david.h.meyer3.civ@army.mil}
\author{Joshua C. Hill}
\author{Paul D. Kunz}
\author{Kevin C. Cox}
\affiliation{%
 DEVCOM Army Research Laboratory, Adelphi MD 20783, USA
}%

\date{\today}% It is always \today, today,

\begin{abstract}
%Quantum 
Electric field sensors based on Rydberg atoms offer unique capabilities, relative to traditional sensors, for detecting radio-frequency (rf) signals.  In this work, we demonstrate simultaneous demodulation and detection of five rf tones spanning nearly two decades (6 octaves), from 1.7\,GHz to 116\,GHz.  We show continuous recovery of the phase and amplitude of each tone and report on the system's sensitivity and bandwidth capabilities for multi-band detection.  We use these capabilities to demonstrate a digital communication protocol, simultaneously receiving on-off-keyed binary data from four bands spanning nearly one decade of frequency.  
\end{abstract}

\maketitle

%\tableofcontents

\section{Introduction}

The radio-frequency (rf) spectrum is typically defined as electromagnetic waves with frequencies between 30\,kHz and 300\,GHz, a range of six decades.  Humans use this entire range for applications such as communication, remote sensing, navigation, and timekeeping.
There is increasing interest in rf sensors based on Rydberg states of atoms, with a highly-excited electron, because they exhibit sensitivity and resonances across the entire range \cite{sedlacek_microwave_2012,adams_rydberg_2019,holloway_electric_2017}. 

In contrast, traditional rf sensors based on resonant antennas, typically have limited bandwidth, operating over a fraction of a decade.  Using multiple rf bands usually requires more than one antenna, increasing a system's size and complexity.   Further, using multiple antennas often leads to co-location interference, where each antenna interferes with the performance of its neighbor.  Rydberg atoms offer a unique path to overcome these technological challenges. 

Recent work in the field of Rydberg sensing has pointed to the capability of Rydberg atoms to detect rf fields across the entire spectrum, using various methods including multiple atomic species \cite{holloway_multiple-band_2021}, Stark tuning of resonances \cite{simons_continuous_2021}, utilizing the off-resonant Stark effect \cite{paradis_atomic_2019,jau_vapor-cell-based_2020,meyer_waveguide-coupled_2021}, multi-photon resonances \cite{anderson_optical_2016}, and operating across the vast forest of Rydberg resonances \cite{meyer_assessment_2020}, with demonstrations ranging from THz \cite{downes_full-field_2020} to quasi-DC \cite{cox_quantum-limited_2018,jau_vapor-cell-based_2020}. We extend upon this work by demonstrating simultaneous multi-band demodulation using multiple Rydberg states.  

First we introduce how the square-law sensitivity of Rydberg atoms in the off-resonant regime allows simultaneous demodulation.  Second, we introduce the experimental apparatus and demonstrate demodulation and detection of five tones, ranging from 1.7\,GHz to 116\,GHz.  We show that the amplitude and phase of each tone can be recovered simultaneously, by demodulating into different portions of the base band, with approximately 6\,MHz of instantaneous bandwidth.  Finally, we demonstrate simultaneous detection of binary digital signals with four tones at bit rates of up to \SI{40}{\kilo\bit\per\second} with bit error rates better than \SI{0.6}{\kilo\bit\per\second}.

\section{Theory Background}

\begin{figure*}[tb]
    \centering
    \includegraphics[width=\textwidth]{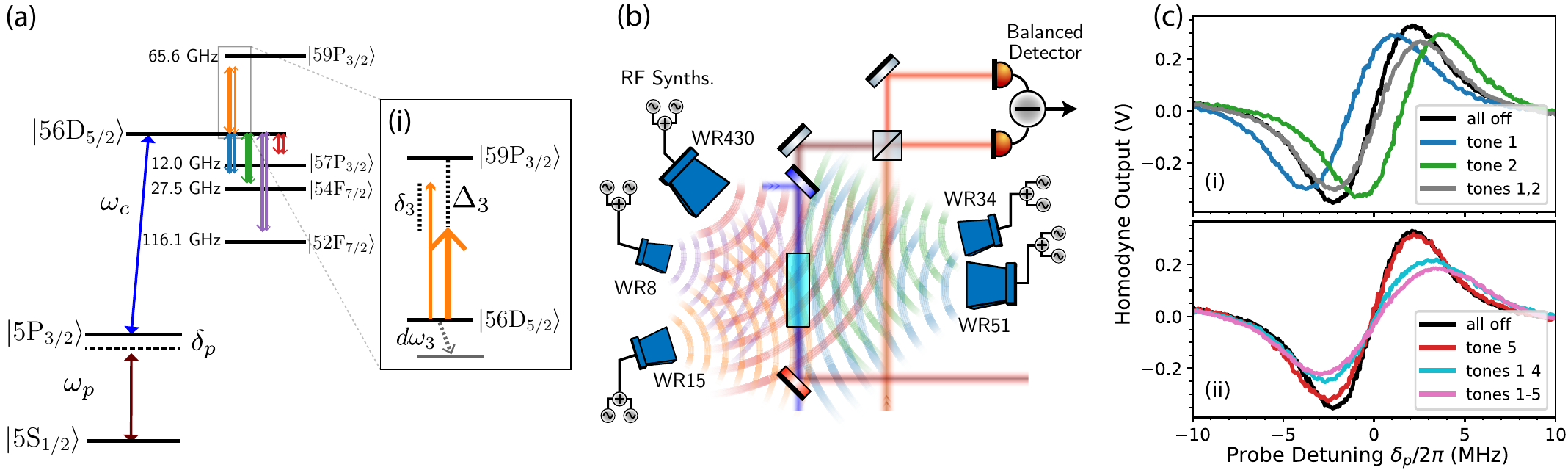}
    \caption{Energy diagrams and experimental configuration.
    (a) Energy level Diagram of the full system. Optical spectroscopy is done in a ladder-EIT configuration (denoted by the maroon and blue arrows). The microwave couplings are shown as arrows branching from the $\state{56}{D}{5/2}$ state. The resonant frequencies of the transitions to nearby Rydberg states are shown.
    (a)(i) Energy level diagram for off-resonant heterodyne measurements using Rydberg states. A strong LO field, detuned from atomic resonance, produces a large Stark shift $d\omega_i$ of the target Rydberg state. A weak signal tone, detuned from the LO field, produces beats in the Stark shift of the target state that can be measured spectroscopically.
    (b) Simplified diagram of the experimental apparatus. The horns corresponding to tones 1 through 5 are fed by waveguides of size WR51, WR34, WR15, WR8, and WR430, respectively.
    (c)(i) Example spectroscopy of the target Rydberg state with and without applied rf with near-detuned couplings. The black trace is the unshifted peak. The blue and green traces correspond to LO 1 or 2 being on, respectively. The gray trace shows both LOs on.
    (c)(ii) Effect of far-detuned couplings. The red trace shows the observed Stark shifts from LO 5. The cyan trace shows the resulting peak with LOs 1-4 on. The magenta trace shows LOs 1-5 on.
    \label{fig:leveldiagram}
    }
\end{figure*}

The Rydberg sensor level diagram is shown in Fig.~\ref{fig:leveldiagram}(a).  We operate the sensor by performing spectroscopy on a single Rydberg ``target'' state.  Interaction with multiple rf tones (denoted by the colored arrows) leads to simultaneous perturbations in the state energy. These perturbations are measured spectroscopically via a ladder-EIT scheme comprised of a probe and a coupling laser.

In order to simultaneously detect the perturbations on the target state from multiple rf signals, it is important that the signal perturbations linearly combine in the spectroscopic response, allowing for a linear recovery of each signal tone. This condition can be stated as
\begin{equation}\label{eq:linSum}
    d\omega = \sum_i d\omega_i
\end{equation}
where $d\omega$ is the total frequency shift of the target Rydberg state and $d\omega_i$ is the frequency shift due to a single rf field $E_i$. If this condition is not satisfied, the spectroscopic response to multiple rf fields is nonlinear; where simultaneous signal recovery generally becomes  complex and inter-dependent. In this work, we employ rf couplings to the target state that satisfy Eq.~\ref{eq:linSum}.

Our sensor simultaneously detects five signals, with frequencies 1.72, 12.11, 27.42, 65.11, and 115.75\,GHz.  The highest four of these signals operate somewhat detuned from nearby Rydberg transitions, while the lowest frequency of 1.72\,GHz, is far red-detuned from all relevant transitions. In this off-resonant regime, the Rydberg energy shifts scale quadratically with the field amplitude
\begin{equation}\label{eq:starkShift}
    d\omega_i = \langle a_i (E_i)^2 \rangle
\end{equation}
where $a_i$ is the polarizability and the angle brackets denote a time average representing the atomic sensor's finite instantaneous response bandwidth. Working in this detuned regime, in contrast to multiple resonant rf couplings, allows for linear addition of the $d\omega_i$ shifts. See Appendix \ref{app:2ToneResp} for further details.

In each band, we apply two fields to implement an rf heterodyne measurement, one signal electric field $S_i$ and one local oscillator $E_i\gg S_i$, as was demonstrated using resonant rf couplings \cite{simons_rydberg_2019,jing_atomic_2020}.
When the signal and local oscillator fields are far detuned from resonance
the resulting state shift is \cite{meyer_waveguide-coupled_2021}
\begin{equation}\label{eq:het}
    d\omega_i = a_i\left[\frac{E_i^2}{2} + \frac{S_i^2}{2}+ E_i S_i \cos\left(\delta_i t \right)\right].
\end{equation}
The first two terms represent static shifts due to the rms amplitudes of the LO and signal fields.
The final term in Eq.~\ref{eq:het} represents the signal field demodulated into the sensor baseband, with baseband frequency $\delta_i$ \footnote{This and other frequency definitions in this work are defined in units of \si{\radian\per\second}. However, we will often specify them in \si{\hertz}. So a beat frequency of \SI{100}{\hertz} would be interpreted as $\delta=2\pi\times\SI{100}{\hertz}$.}. Due to the square-law response to the applied field when working off resonance, this term also has heterodyne gain from the amplitude of the LO field itself. 

Electromagnetically-induced transparency (EIT) spectroscopy is used to detect the shifted resonance frequency of the target state $d\omega$, which contains information from all of the applied fields $E_i+S_i$ according to Equations \ref{eq:linSum} and \ref{eq:het}.

\section{Experimental Setup}

The experimental layout and atomic level diagrams are shown in Fig.~\ref{fig:leveldiagram}(a). Using two lasers (denoted as ``probe'' (dark red) and ``coupling'' (blue)), we perform EIT spectroscopy on the $\ket{56D_{5/2}}$ target state to simultaneously detect five rf signal tones, colored orange, blue, green purple, and red.  The Rydberg demodulation scheme for a single tone is shown in the inset (Fig.~\ref{fig:leveldiagram}(a)(i)).  For each tone $i$, a strong local oscillator field $E_i$ is applied from a microwave horn, with detuning from the target state $\Delta_i$. The corresponding signal field $S_i$ has a small detuning $\delta_i$ from the LO, and is broadcast from the same corresponding horn to super-impose it with $E_i$ at the atomic sample.  By Eq.~\ref{eq:het}, this leads to a detected baseband signal (a heterodyne beat) with frequency $\delta_i$.  

Figure \ref{fig:leveldiagram}(b) depicts the experimental setup.  
Five separate rf horns are aimed at the Rb vapor cell from approximately 30\,cm away. The cell is 7.5\,cm long and contains natural abundance Rb. A narrow transmission resonance (EIT) is observed when the probe laser detuning $\delta_p$ is swept across the $\ket{5S_{1/2}}\rightarrow\ket{56D_{5/2}}$ two-photon resonance. These spectroscopy signals are detected using the balanced optical homodyne configuration described in \cite{meyer_assessment_2020}, operated in the phase quadrature, yielding a dispersive-shaped EIT feature as the probe laser detuning is swept.
To avoid ambiguity with the rf signals, we will refer to the output homodyne signal as the ``output'' for the remainder of this manuscript.
For all data shown, except Fig.~\ref{fig:leveldiagram}(c), the probe is locked to the center of the EIT dispersive feature where the slope (and therefore sensitivity) is largest.

The frequencies of the five rf tones, their index labels, and their respective field amplitudes at the atomic sample are shown in Tab.~\ref{tab:tones}.
Tones 1 through 4 are nearly resonant to a dipole-allowed Rydberg transition between the target state and the nearby state indicated in the table. Tone 5 is far red-detuned ($>\SI{10}{\giga\hertz}$) from all relevant Rydberg transitions.
The inclusion of tone 5 in this demonstration highlights the broad operational range of the protocol relative to Rydberg transitions and that multiple tones can interact with the same Rydberg transition.
The magnitudes of the signal fields are chosen such that the beat of each tone has approximately the same signal-to-noise ratio; differences in set-point primarily reflect differences in the Rydberg sensitivity at that frequency via changes to the atomic polarizability $a_i$.
This sensitivity, and linear dynamic range, depend on the specific sensor configuration, including: detuning from Rydberg resonances and rf LO powers and frequencies.

The LO and signal fields for tones 1, 2, and 5 are generated directly from commercial signal generators. The LO and signal fields for tones 3 and 4 are produced using synthesizers that are then multiplied by a factor of 6 before being broadcast onto the atoms. Further details of the setup are given in App.~\ref{app:ExpDetails}.

In Fig.~\ref{fig:leveldiagram}(c) we show representative phase quadrature optical EIT output with various combinations of LO fields applied to the atoms.
In Fig.~\ref{fig:leveldiagram}(c)(i) we show example probe sweep spectroscopy traces for LOs 1 and 2, which are relatively close detuned from Rydberg resonances. The black trace shows the unshifted dispersive resonance. The blue and green traces correspond to LO 1 or 2 being on, respectively, resulting in Stark shifts of the resonance in opposing directions. The gray trace shows both LOs on, resulting in nominal canceling of the Stark shifts, though the resonance is broadened.
This broadening is predominantly the result of populating multiple $m_J$ sublevels of the target state, where each sublevel results in a different dipole moment for the rf field. This specific source of broadening could be mitigated by only populating a single sublevel of the target state.
Other potential sources of broadening, such as power broadening or inhomogeneous fields, would manifest similarly and should also be considered when optimizing sensor performance.

\begin{table}[tb]
\begin{tabular}{c w{c}{4em} S S[table-column-width=4.5em] w{c}{2.5em} S S} \toprule 
Tone  & State         & {$\omega_0^i$}  & {$\Delta_i$} & {$\delta_i$} &  {$E_i$} & {$S_i$} \\
$i$  & $\state{n}{L}{J}$ & \multicolumn{2}{c}{\si{\giga\hertz}} & {\si{\kilo\hertz}} & \multicolumn{2}{c}{\si{\volt\per\meter}} \\
\cmidrule(r){1-2}\cmidrule(r){3-5}\cmidrule{6-7}
1       & $\state{57}{P}{3/2}$     & 12.01         & +0.1 &  75 & \num{1.2(1)} & \num{0.12(1)}\\
2       & $\state{54}{F}{7/2}$     & 27.52         & -0.1 & 130 &  \num{2.4(2)} & \num{0.14(1)} \\
3       & $\state{59}{P}{3/2}$     & 65.61         & -0.5 & 180 &  \num{17.9(26)} & \num{1.5(3)}\\
4       & $\state{52}{F}{7/2}$     & 116.05        & -0.3 & 240 &  \num{36.0(36)} & \num{1.4(2)}\\
5       & $\state{57}{P}{3/2}$     & 12.01         & -10.287 & 280 &  \num{16.0(4)} & \num{2.3(1)}\\
\bottomrule
\end{tabular}
\caption{Rf tone experimental parameters, including the tone number, nearby Rydberg state being coupled to, resonance frequency of that coupling $\omega_0^i$, detuning of the LO field relative to that resonance $\Delta_i$, typical signal detuning from the LO field $\delta_i$ unless otherwise noted, and the signal and LO field amplitudes at the atomic sample $E_i$ and $S_i$, as described in App.~\ref{app:FieldCal}.}
\label{tab:tones}
\end{table}

In Fig.~\ref{fig:leveldiagram}(c)(ii) we show the effect of far-detuned couplings. The red trace shows the observed Stark shift from LO 5. Due to the large detuning, there is no common shift in one direction as the sublevels of the $\text{D}_{5/2}$ state do not all shift in the same direction (see App.~\ref{app:FieldCal} for further details). The cyan trace shows the dispersive resonance with LOs 1--4 on, resulting in a largely canceled shift, but broader feature. The magenta trace shows LOs 1--5 on, which leads to even more broadening of the dispersive feature.
Since the sensitivity of the sensor is related to the slope of this dispersive feature (which converts Stark shifts to spectroscopic changes in the measured homodyne probe phase), broadening from increasing numbers of LO tones limits the number of signals that can be simultaneously measured with a given sensitivity.

\section{Results}

\subsection{Simultaneous Tone measurement}

\begin{figure}[tb]
    \centering
    \includegraphics[width=\columnwidth]{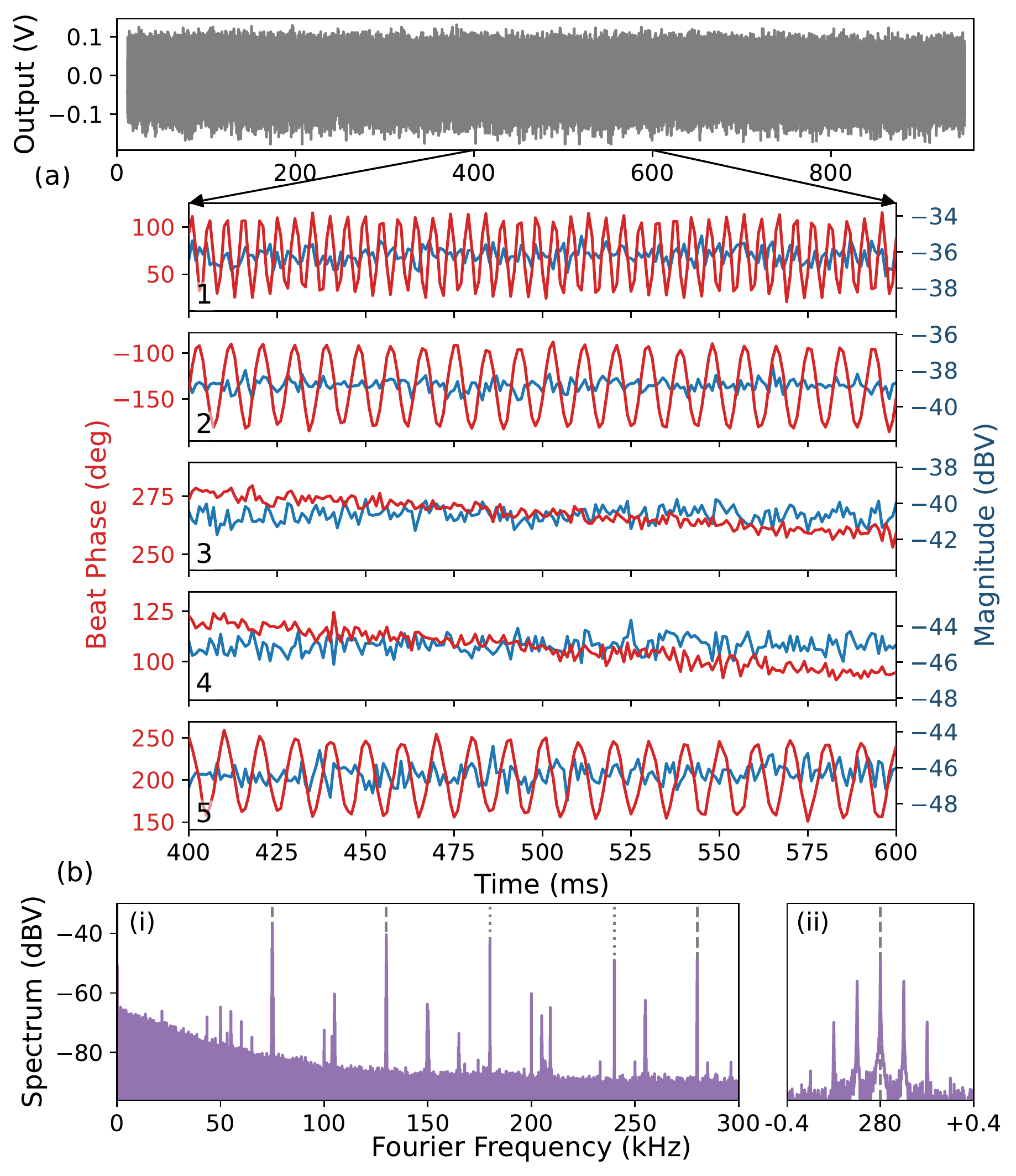}
    \caption{Simultaneous amplitude and phase recovery of tones spanning over 114\,GHz.
    (a) Single timetrace of the photodetector output acquired at 2\,MS/s over 1\,s.
    (a)1-5 correspond to the amplitude (blue) and phase (red) recovery for tones 1-5, respectively. Tones 1, 2, and 5 had slow phase modulations applied to the signal field of \SI{205}{\hertz}, \SI{110}{\hertz}, and \SI{100}{\hertz} respectively. 
    (b)(i) FFT of the full timetrace in part (a). Resolution of the FFT is 1\,Hz.
    (b)(ii) Zoomed view of the FFT around tone 5 at \SI{280}{\kilo\hertz}, showing the clearly resolved phase modulation sidebands.}
    \label{fig:timetrace}
\end{figure}

We first present data showing simultaneous demodulation of five tones into the baseband.
We broadcast tones 1-5 onto the atoms and record a time-series of the homodyne output for \SI{1}{\second}.
The raw output is displayed at the top of Fig.~\ref{fig:timetrace}(a), and the processed data reveal the tones received from all five bands.
The experimental settings are summarized in Tab.~\ref{tab:tones}.
In these data, tones 1, 2, and 5 have \SI{0.8}{\radian} sinusoidal phase modulations applied with rates of 205\,Hz, 110\,Hz, and 100\,Hz respectively.
All five tones are distinctly visible in the corresponding power spectrum (Fig. \ref{fig:timetrace}(b)(i)), found by taking the Fourier transform of the timetrace in part (a).
A enlarged subset of the power spectrum showing the detailed response in the vicinity of tone 5 is shown in Fig.~\ref{fig:timetrace}(b)(ii), where the phase modulation sidebands are evident.
We identify the other beats in the spectrum as harmonics and sum or difference frequencies of the five fundamental beats.
The spurious tones are reduced by more than \SI{20}{\decibel} from the fundamental tones and are indicative of a weakly nonlinear process which we attribute to residual nonlinearity in the optical homodyne detection. Higher-order effects in the atomic system not considered in this work may also be contributing.
We independently identified three spurious tones due to electronic pickup at 174, 180, and \SI{209}{\kilo\hertz}.

We simultaneously recover the phase and amplitude of each individual tone via suitable post-processing of the timetrace.
We take the Fourier transform of the raw data, and extract each tone in frequency space with a \SI{1}{\kilo\hertz} bandwidth.
The amplitude and phase correspond to the magnitude and angle of the complex Fourier component.
The phase and amplitude of each beat tone is displayed in Fig.~\ref{fig:timetrace}(a)(1-5), shown as the red and blue traces respectively.
The amplitudes of each tone are constant, with fluctuations in the recovered amplitude consistent with the Rydberg sensor noise floor observed in Fig.~\ref{fig:timetrace}(b).
This noise floor is largely white with $1/f$ probe laser phase noise dominating at frequencies below approximately \SI{100}{\kilo\hertz} because of unequal interferometer path lengths in the homodyne detection which results in uncancelled laser phase noise.
The overall phase offset of each beat corresponds to the uncontrolled phase offset of the LO and signal frequency sources between each tone.
The phase modulations of tones 1, 2, and 5 manifest as sinusoidal variations of the recovered phase, matching the phase deviation (\SI{90}{\degree} peak-to-peak) and modulation rates applied.
The synthesizers for tones 3 and 4 are not modulated.
The observed drifts in recovered phase instead correspond to the relative phase drift between the LO and signal channels for each synthesizer.

\begin{figure}[tb]
    \centering
    \includegraphics[width=\columnwidth]{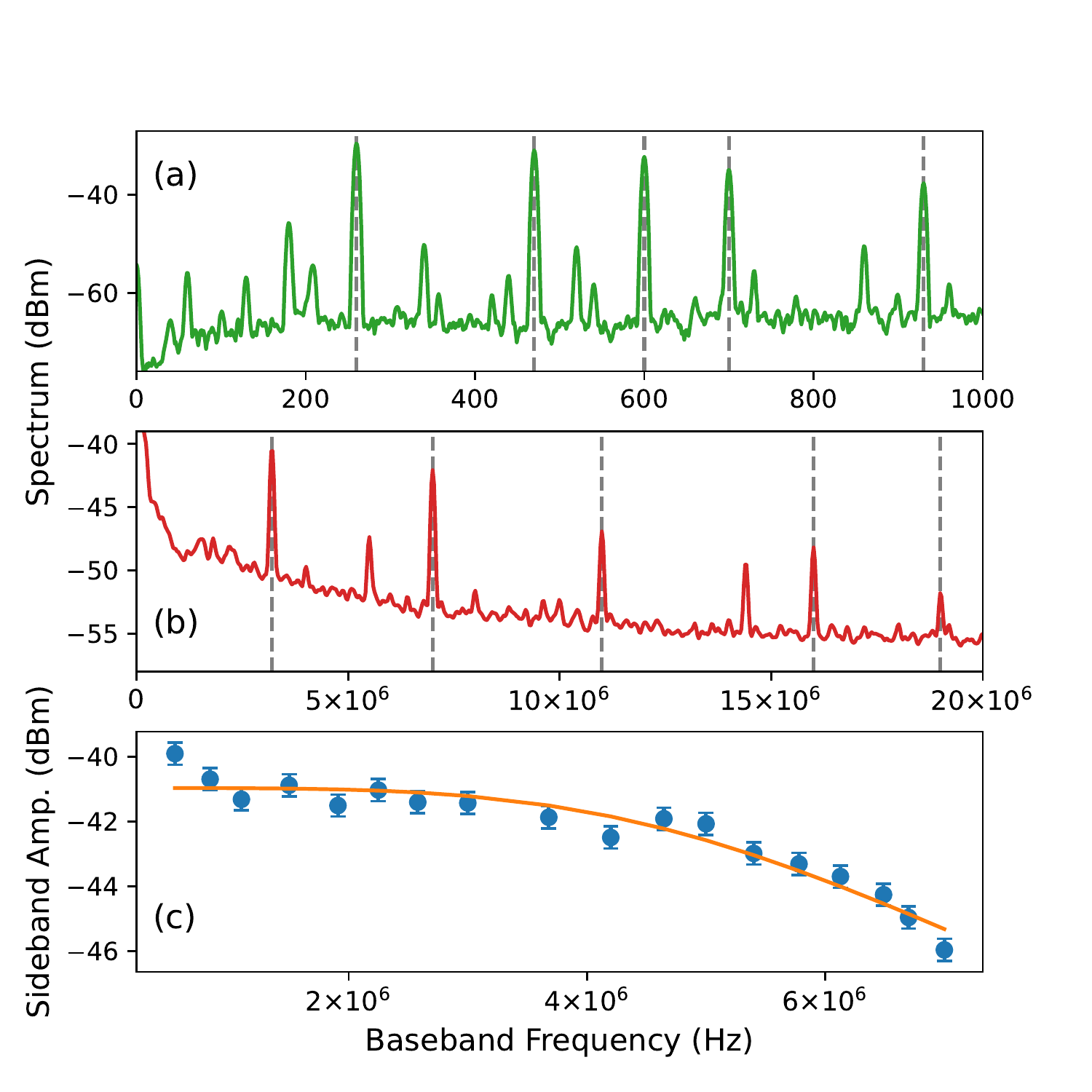}
    \caption{Spectrum analyzer measurements of the EIT output with five rf heterodynes applied. Vertical gray dashed lines denote the frequency of the applied beat tones. Each spectrum is an average of 20 traces.
    (a) Beats at 260, 470, 600, 700, and \SI{930}{\hertz} for tones 1-5. Resolution bandwidth is 4.89\,Hz.
    Note that the $1/f$ noise present in other spectra is suppressed in this frequency range due to active optical homodyne path stabilization.
    (b) Beats at 16, 19, 11, 7, and \SI{3.2}{\mega\hertz} for tones 1-5, respectively. While the atomic response reduces beat sizes at higher frequencies, they are still visible. The photodetector frequency response is removed from this dataset. Resolution bandwidth is 97.75\,kHz.
    (c) Output beat amplitude versus modulation frequency. Phase modulation of 0.8\,rad is applied to tone 2 (beat frequency at \SI{130}{\kilo\hertz}) at varying modulation frequencies. The orange line is a fit to a second-order low-pass function with \SI{3}{\decibel} corner at \SI{6.11(16)}{\mega\hertz}. The photodetector response is removed from the data (see App.~\ref{app:DetCal}).}
    \label{fig:PM-SA}
\end{figure}

The total capacity for the Rydberg sensor to detect multiple tones is governed by the baseband instantaneous bandwidth, which is primarily set by the optical spectroscopy method \cite{meyer_digital_2018}.
This measured bandwidth is shown in Fig.~\ref{fig:PM-SA}.
Part (a) and (b) are representative five-tone power spectra with different base-band detunings $\delta_i$.
In part (a) all $\delta_i < 2\pi\times\SI{1}{\kilo\hertz}$ and in part (b) all $\delta_i$ are distributed over \SI{20}{\mega\hertz}.
The output reduction due to the instantaneous bandwidth rolloff is evident in part (b).
In Fig.~\ref{fig:PM-SA}(c), we directly measure the bandwidth by transmitting rf with a Fourier component of fixed amplitude and varying frequency.
We accomplish this by varying the frequency of a phase modulation on tone $i=2$.
We fit the measured data to a double-pole low-pass filter (shown as the orange line in Fig.~\ref{fig:PM-SA}(c)) that has a \SI{3}{\decibel} point of \SI{6.11(16)}{\mega\hertz}.
The number of individual beat tones that can be simultaneously measured depends on this bandwidth and the resolution bandwidth of the measurement.  For example, using 10\,kHz spacing between output beats as is done here, our 6.1\,MHz bandwidth could nominally allow for 610 tones.

Applying many individual rf heterodyne pairs may be challenging due to the increased broadening of the Rydberg resonance with each LO and the potential for intermodulations between the beats, described above. It is possible to apply multiple signal beats to a single LO, helping to mitigate this limitation.  In any case, the demonstration highlights the ability of the atoms to demodulate many highly disparate frequency tones (spanning 1.7 to 116\,GHz) into a single base band, detected optically.

\subsection{Communications}

\begin{figure}[tb]
    \centering
    \includegraphics[width=\columnwidth]{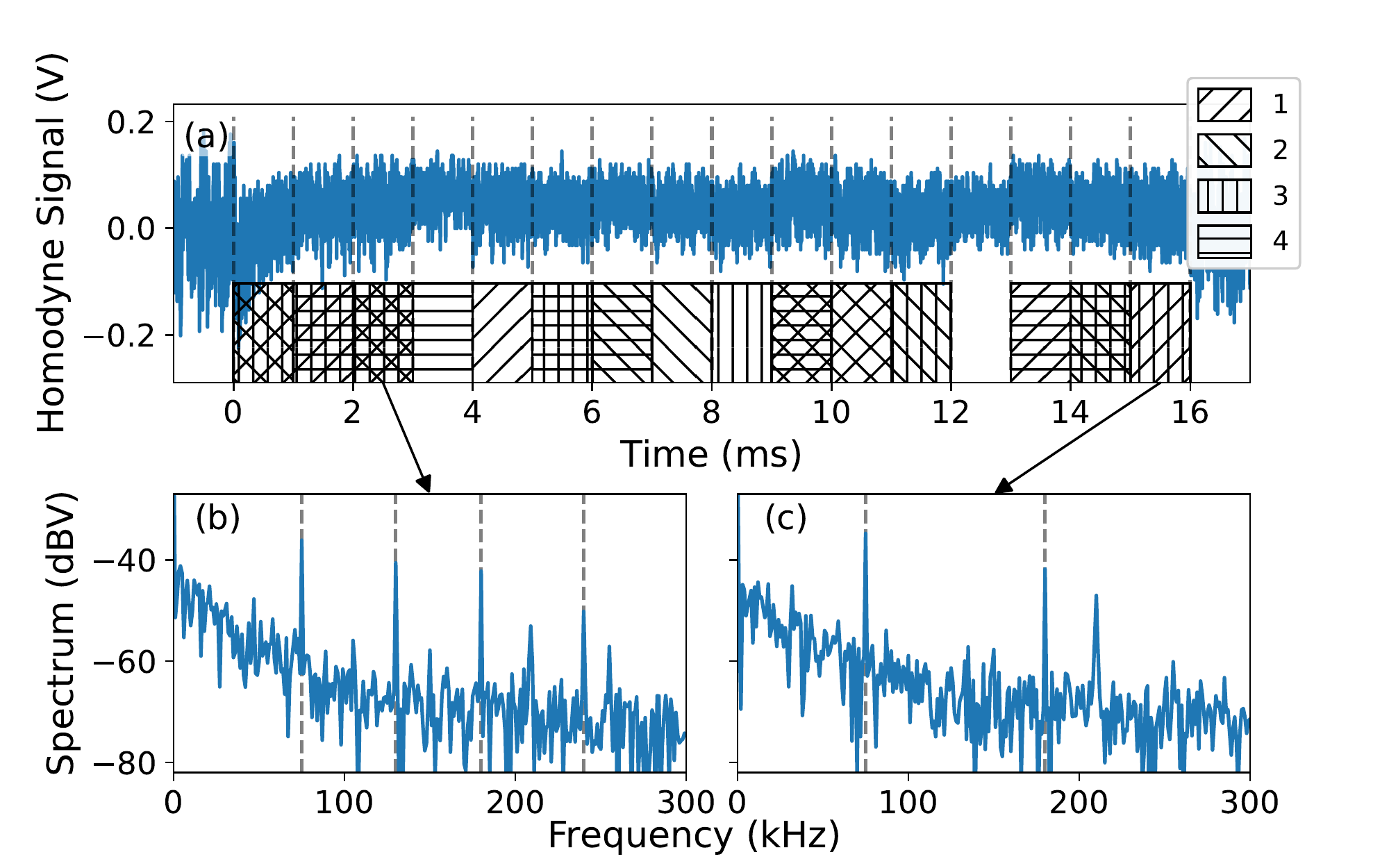}
    \caption{Example symbol recovery of a communications signal.
    (a) Shows the raw timetrace of the EIT output from the photodetector, taken with a sample rate of \SI{4}{\mega\sample\per\second}. The hashed boxes along the bottom axis denote which beat tones are on for each symbol period of \SI{1}{\milli\second}: tone 1 is the right diagonal, tone 2 is the left diagonal, tone 3 is the vertical, and tone 4 is the horizontal.
    (b) Shows the FFT of the third symbol in the sequence. This symbol has all four tones on, their locations noted by gray-dashed vertical lines.
    (c) Shows the FFT of the 15th symbol in the sequence. This symbol only has tones 1 and 3 on.}
    \label{fig:rawcomms}
\end{figure}

We perform four-band communication reception using an on-off keying (OOK) protocol with tones 1 through 4 as described in the previous section, though far-off-resonant tones such as tone 5 could also be used. 
The OOK protocol does not encode information in the phase of the beats, however the Rydberg sensor can observe the phase of each beat and the relative phase between beats.
Protocols that encode information in these relative phases are interesting to consider, but require transmitters to be phase-synchronous.
The four tones create 16 individual digital symbols, where an off tone represents a zero and an on tone represents a one.  We step through these symbols, sending each for a fixed period of time, by turning the signal sources on and off (the LOs stay on for the duration). 
An example is shown in Figure \ref{fig:rawcomms}, where each symbol is sent for 1\,ms. Part (a) is the raw timetrace, with gray dashed vertical lines separating each symbol's time window and the hashed boxes denote which signal tones are on for each.
We recover the symbol that is present by taking a Fourier transform in each window and comparing the amplitude at each beat frequency with a known, fixed background level. Figure~\ref{fig:rawcomms}(b-c) shows spectra for time windows 3 and 15. Window 3 has all tones on, and window 15 has only tones 1 and 3 on.

We measure the fidelity of symbol recovery under different experimental conditions by sending randomized symbol sequences and comparing the recovered OOK data stream with the known sent patterns.
Table~\ref{tab:BER} displays the directly-measured bit error rates for the various settings, that include higher or lower signal fields (representing a difference of \SI{10}{\decibel} in signal power) and symbol periods of 1\,ms and \SI{100}{\micro\second}.
For the slower symbol rate, representing a data rate of \SI{4}{\kilo\bit\per\second}, the bit error rate is better than \SI{8e-3}{\kilo\bit\per\second}.
The faster symbol rate, yielding a data rate of \SI{40}{\kilo\bit\per\second}, has bit error rates better than \SI{6.1e-1}{\kilo\bit\per\second}.

\begin{table}[tb]
\begin{tabular}{ccccS[table-column-width=5.5em]} \toprule
     Sig.\,Fields & Period & BR & BE & {BER} \\
     \si{\volt\per\meter} & \si{\milli\second} & \si{\kilo\bit\per\second} & \% & {\si{\kilo\bit\per\second}}\\
     \cmidrule[0.5pt](){1-3}\cmidrule[0.5pt](l){4-5}
     \multirow{2}{9em}{0.11, 0.13, 0.93, 2.0} & 1 & 4 & 0.2 & \num{7.8e-3} \\
      & 0.1 & 40 & 1.5 & \num{6.1e-1} \\ \midrule[0.1pt]
     \multirow{2}{9em}{0.36, 0.41, 2.9, 6.3} & 1 & 4 & 0.05 & \num{1.9e-3} \\
      & 0.1 & 40 & 0.01 & \num{2.8e-3} \\\bottomrule
\end{tabular}
\caption{Bit error summary of the OOK scheme under different signal amplitudes and symbol periods. The signal field amplitudes are listed in order for $i = 1\text{--}4$. The bit-rate (BR) is the corresponding data rate. The bit error ratio (BE) shows the fraction of bits that were incorrectly received and the bit error rate (BER) is the number of bit errors per unit time.}
\label{tab:BER}
\end{table}

This communication reception demonstration is conceptually simple. However, it highlights the power of simultaneous data reception of multiple signal carriers spanning from 12 to 116\,GHz. Given the broader capability of the Rydberg sensor, which includes full amplitude or phase demodulation of many tones (potentially $>100$), it is possible to implement full quadrature-amplitude modulated encoding for each signal tone which would represent a complete orthogonal frequency domain multiplexing protocol that can span multiple decades of the rf spectrum \cite{hanzo_single-_2000}.

\section{Conclusion}

We have shown how to utilize the multi-resonant response profile of a Rydberg atomic sensor to detect a wide range of rf tones.
This intrinsic multi-band nature of atoms is likely to lead to additional sensing capabilities in the future, where it is difficult to achieve a similar ultra-wide response profile using a sensor with a single resonant antenna.
While the goal of this demonstration is not state-of-the-art sensitivity for Rydberg sensors, it shows a capability of such devices that cannot be easily mimicked by other physical platforms.

A primary remaining challenge is to improve Rydberg sensors' sensitivity to weak fields. The fields detected in this work, for example, are of order $10^4$ times stronger than those detected for GPS triangulation.  However, this limitation is governed by constraints of the EIT-based spectroscopy method \cite{meyer_optimal_2021} and is not a limit inherent to atomic or quantum sensors. With further research, improved sensitivities are possible \cite{fan_atom_2015,adams_rydberg_2019}, yielding high performance receivers that span a large section of the spectrum.

Another important avenue for study is increasing the instantaneous bandwidth of the Rydberg sensor.  The current bandwidth roll-off near 6.1\,MHz is governed by the EIT-based spectroscopy method.  Alternate sensing schemes may improve this bandwidth, and therefore the multiplexing capacity, of Rydberg sensors used for ultra-wideband detection.  

\section*{Acknowledgments}

The authors recognize financial support from the Defense Advanced Research Projects Agency (DARPA) Quantum Apertures program.

The views, opinions and/or findings expressed are those of the authors and should not be interpreted as representing the official views or policies of the Department of Defense or the U.S.~Government.

References to commercial devices do not constitute an endorsement by the US Government or the Army Research Laboratory. They are provided in the interest of completeness and reproducibility.

\appendix

\section{Resonant vs Off-resonant Two-Tone Response}
\label{app:2ToneResp}

We show that off-resonant rf couplings to the target Rydberg state are generally necessary to satisfy the conditions of Eq.~\ref{eq:linSum}. We begin by defining an interaction Hamiltonian of our optically-probed target Rydberg state coupled to two Rydberg states via rf fields with Rabi frequencies $\Omega_a$, $\Omega_b$ and detunings $\delta_a$, $\delta_b$.
\begin{equation}
    H = \frac{1}{2}\begin{pmatrix}
        0 & \Omega_a & \Omega_b\\
        \Omega_a & 2\delta_a & 0\\
        \Omega_b & 0 & 2\delta_b
        \end{pmatrix}
\end{equation}
We then determine the eigenfrequencies by diagonalizing this Hamiltonian.
These eigenfrequencies represent the shifts of the coupled eigenstate energies from the bare Rydberg target state energy.
It is these resonance shifts that are detected by performing EIT spectroscopy of the Rydberg target state (see Fig.~\ref{fig:leveldiagram}(a)(i)). By analyzing the eigenfrequencies under different rf detunings, we can illustrate how multiple rf couplings to the same Rydberg state influence the observed spectroscopic shifts.

If both rf couplings are resonant with Rydberg transitions ($\delta_a$, $\delta_b=0$), the eigenfrequencies are the quadrature sum of both Rabi frequencies.
\begin{equation}
    d\omega_\text{res}=\pm \frac{1}{2}\sqrt{\Omega_a^2+\Omega_b^2}
\end{equation}
This quadrature sum prevents general linear mapping of the individual couplings into the sensor's baseband.
While signal recovery is still possible with such a nonlinear response, the coupled shifts often lead to undesirable results. For example, if $\Omega_a\gg\Omega_b$, the shift for field $b$ (and any corresponding beat signal) is reduced by an additional factor of $\Omega_b/2\Omega_a$ and any beat in field $b$ will not be independent of beats in field $a$.

If both rf couplings are off-resonant with Rydberg transitions ($\delta_a$, $\delta_b\neq0$), the resulting eigenfrequencies show that the leading-order Stark shifts from each transition sum. If we impose the simplifying assumption that $\delta_i\gg\Omega_i$ and expand to lowest orders in $\Omega_i$, the Stark shift of the target state is
\begin{equation}
    d\omega_\text{off-res}=-\frac{\Omega_a^2}{4\delta_a}-\frac{\Omega_b^2}{4\delta_b}
    +\frac{\Omega_a^2\Omega_b^2}{16\delta_a\delta_b^2}
    +\frac{\Omega_a^2\Omega_b^2}{16\delta_a^2\delta_b}
\end{equation}
As expected, we observe that each coupling's total Stark shift adds linearly into the sensor's baseband. By employing rf heterodyne detection for each tone, the signal response at each coupling is linearized, allowing for linear mapping of each signal tone into the sensor baseband. Note that changing the signs of the detunings also allows for cancelling large dc shifts due to the LOs.

If one rf coupling is resonant and the other is detuned ($\delta_a\neq0$, $\delta_b=0$), one still observes linearly adding shifts from the two couplings. If we again impose that $\delta_a\gg\Omega_a$ and expand to lowest orders in $\Omega_a$, we observe the new eigenfrequencies of the Hamiltonian to be,
\begin{equation}
    d\omega_\text{mixed}=\pm\frac{\Omega_b}{2}-\frac{\Omega_a^2}{8\delta_a}
\end{equation}
In essence, one observes a combination of Autler-Townes splitting and an ac Stark shift. Combined with the above derivations, this shows that Stark shifts can satisfy Eq.~\ref{eq:linSum}, so long as no more than one rf coupling is resonant. In future applications, this mode of operation could be used to increase the sensitivity of a single tone (by working on resonance), while maintaining wideband demodulation into the sensor baseband for many off-resonant couplings.

\section{Experimental Details}
\label{app:ExpDetails}

The details of the experimental apparatus are described in \cite{meyer_assessment_2020, meyer_waveguide-coupled_2021}.
Here we summarize the most relevant aspects and any deviations from these works.

The homodyne probe field has a power of \SI{3.6}{\micro\watt} and a $1/e^2$ radius of \SI{410}{\micro\meter}. The corresponding Rabi frequency is $\Omega_p = 2\pi\times\SI{3.1}{\mega\hertz}$. The probe laser frequency is stabilized to a spectroscopy-referenced laser.
When the probe laser is not being swept, its frequency is set to be single photon resonant with the atomic transition for zero-velocity class atoms ($\delta_p=0$).
The coupling laser has a power of \SI{500}{\milli\watt} and a $1/e^2$ radius of \SI{380}{\micro\meter}.
The corresponding Rabi frequency, averaged over the $|m_J|=1/2,\, 3/2$ couplings, is $\Omega_c=2\pi\times\SI{4.9}{\mega\hertz}$
The coupling laser frequency is stabilized to a ultra-low-expansion glass optical cavity.
The coupling laser frequency is set to be single photon resonant with the atomic transition for zero-velocity class atoms.

Balanced optical homodyne detection is implemented using a Thorlabs PDB450A detector set to a gain of $10^5$. The homodyne path phase is stabilized using an additional optical tone present on the probe laser, that is detected in heterodyne using a bias tee with both low and high pass port bandwidths of 32\,MHz. The ``DC" (homodyne) output of this bias tee is the source of all data presented in this work.

We acquire data from the homodyne output in three primary ways: USB-DAQ, oscilloscope, and FFT spectrum analyzer. The first data acquisition method is a sweep of the probe beam frequency $\omega_p$ across its single-photon resonance, for a fixed $\omega_c$. We then record the resulting EIT output (e.g. Fig.~\ref{fig:leveldiagram}c) at 500\,kS/s using a National Instruments USB 6343 data acquisition system (USB-DAQ). The second method is recording time domain voltage data using a Keysight DSOX3024T oscilloscope. From this, the data can be presented directly (Fig.~\ref{fig:timetrace}a) or transformed to the frequency domain in post-processing (Fig.~\ref{fig:timetrace}b i and ii) via Fast Fourier Transform (FFT). The third method is acquiring power spectra directly using the FFT spectrum analyzer function of a multipurpose Liquid Instruments Moku:Lab (0-250\,MHz input bandwidth). Where appropriate, we quote resolution bandwidths (RBWs) associated with measurements taken using this instrument.

We broadcast the five microwave tones (Tab.~\ref{tab:tones}) through free space onto the atomic sample. Each tone consists of a pair of frequencies, signal and local oscillator, that are combined prior to emission from the same horn. The horns for tones 1 through 5 have gains of 20, 15, 25, 25, and 10\,dBi, respectively. The frequencies for tones 1, 2, and 5 are generated by pairs of nearly identical frequency synthesizer models; Rohde \& Schwarz SM100B, Keysight 8257N/D, and Stanford Research Systems SG386, respectively. The signal chains generating tones 3 and 4 begin with seed frequencies generated by two Windfreak SynthHD PRO synthesizers, each with two independent channels that we use for the respective tone's signal and LO. We use VDI rf multiplication modules to multiply the seed frequencies by a factor of six and provide amplification. The modules contain an internal voltage controlled attenuator that we augment with manual waveguide attenuators on the output to provide fine power adjustments. The \texttt{labscript} software suite is used for experimental hardware control and timing \cite{starkey_scripted_2013}.

\section{Field Calibration}
\label{app:FieldCal}

\begin{figure}[tb]
    \centering
    \includegraphics[width=0.9\columnwidth]{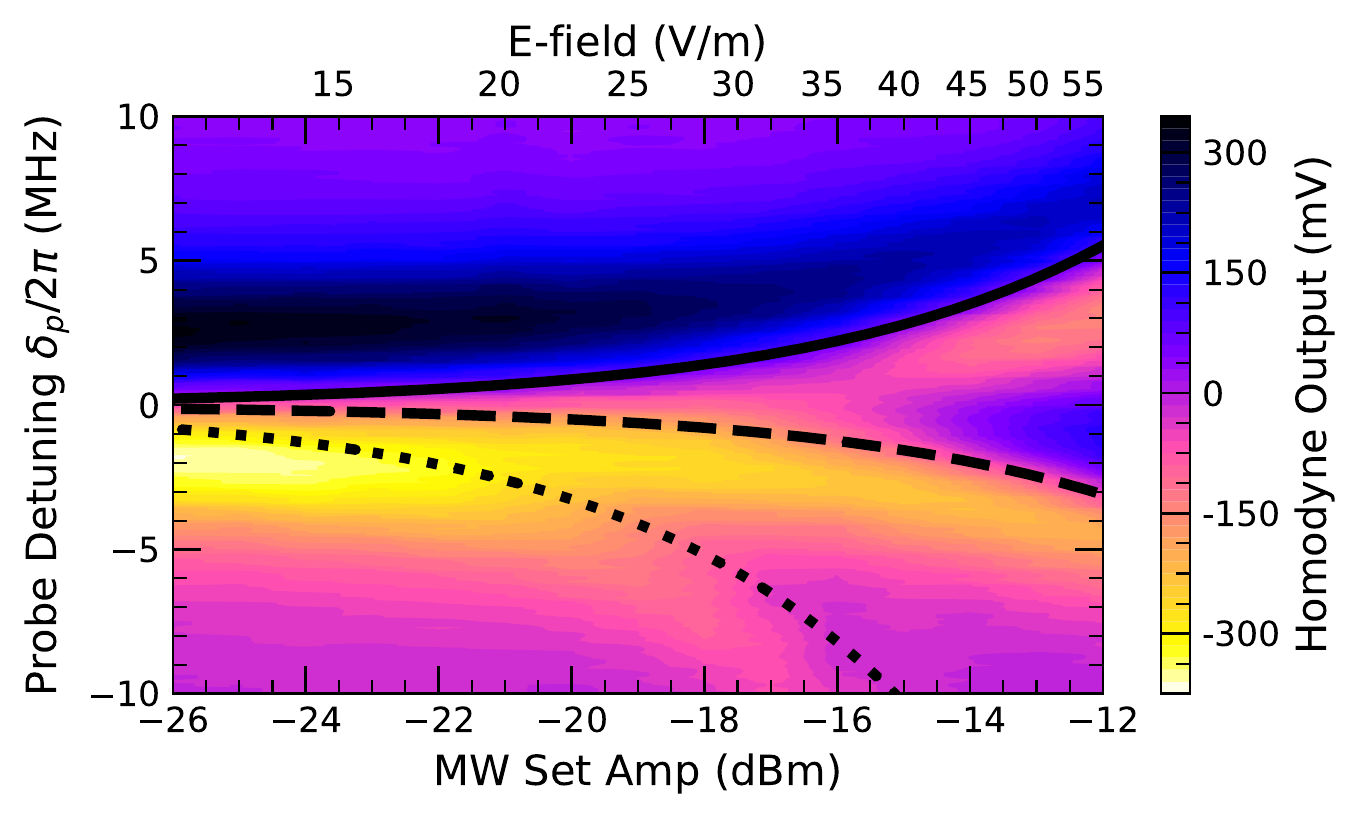}
    \caption{Example calibration of the LO power for tone 5. Due to the large detuning from the nearest Rydberg transition, the atomic sublevels do not shift in the same direction. We match the Floquet theory model to a range of LO rf powers as shown by the solid, dashed, and dotted lines, which track the zero crossing of the dispersive resonance for $|m_J|=0.5$, 1.5, and 2.5, respectively.}
    \label{fig:Tone5Cal}
\end{figure}

We calibrate rf field amplitudes at the atoms using the known spectroscopic response of the Rydberg target state. This response is calculated using the Floquet analysis developed in Ref.~\cite{meyer_assessment_2020}. This model allows for arbitrary rf frequencies and properly accounts for Stark shifts from many Rydberg states simultaneously. 

We first calibrate the LO field amplitudes by measuring the amplitude of the Stark shift $d\omega_i$, with one LO field applied at a time, using a probe sweep measurement like those shown in Fig.~\ref{fig:leveldiagram}(c). We then use the Floquet model to determine the field amplitude necessary to produce that shift, accounting for the reduction in observed shift due to Doppler effects in a probe sweep \cite{meyer_digital_2018}, and averaging the response between the nominally populated $|m_J|=1/2,\,3/2$ sublevels.
Reported uncertainty in the field calibrations is due to statistical uncertainty in measuring $d\omega_i$ for each tone.

For tone 5, the carrier frequency is far enough from resonance that the response of the individual sublevels of the $\state{56}{D}{5/2}$ state is more complicated than a simple global shift. To calibrate its field, we take a series of probe sweeps with the LO power varied around its setpoint, and match the Floquet model to the observed shifts. The result is shown in Fig.~\ref{fig:Tone5Cal}, where the solid, dashed, and dotted lines represent the theory predictions of the Stark shifts for the $|m_J|=1/2,\,3/2$, and 5/2 sublevels. Note that the contrast of the peak for the $|m_J|=5/2$ transition is much lower than the other two sublevels. This indicates that this sublevel is not as strongly populated, which is to be expected due to the linear polarizations of all the optical fields, which preclude direct dipole-allowed coupling into that sublevel from the $\state{5}{P}{3/2}$ intermediate state.
This data and corresponding model fit further indicate that the observed broadening shown in Fig.~\ref{fig:leveldiagram}(c) is largely attributable to the sublevel structure of the Rydberg states we are probing.

\begin{figure}[tb]
    \centering
    \includegraphics[width=0.9\columnwidth]{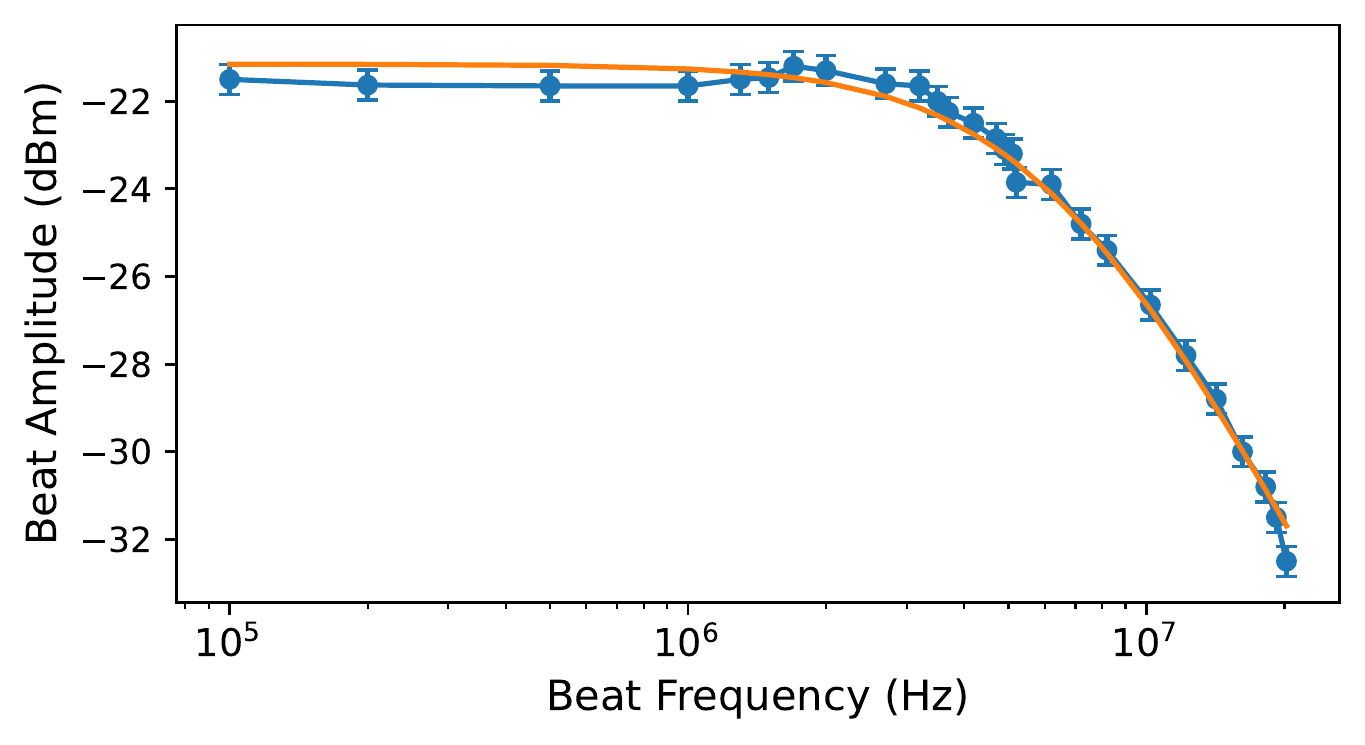}
    \caption{Photodetector response as a function of optical beat frequency. Error bars represent measurement error and the orange line represents a single-pole roll-off fit with a corner at \SI{6.30(14)}{\mega\hertz}.}
    \label{fig:PDBW}
\end{figure}

Calibration of the signal field amplitudes is more difficult, as the weak field amplitudes do not give resolvable Stark shifts on their own. For tones 1, 2, and 5 the signal tone is generated by a signal source identical to the LO signal source, and the two tones are combined using resistive power combiners with power imbalance better than \SI{0.2}{\decibel}. As such, we extrapolate the signal field calibration from the LO field calibration based on the difference in setpoints of the signal generators. For tones 3 and 4, this process is not reliable as the multiplication modules are not guaranteed to amplify the LO and signal tones evenly. To calibrate these tones, we measure beat amplitudes of tones 1 and 2 on a spectrum analyzer, and use the derived LO and signal fields already calibrated to establish a conversion factor between signal field and photodetector output voltage, employing Eq.~\ref{eq:het}. We then use this calibration factor to determine the signal field amplitudes of tone 3 and 4 relative to the calibrated LO field amplitudes.

\section{Detector Bandwidth Correction}
\label{app:DetCal}

The bandwidth response of the sensor is limited, in part, by the photodetector bandwidth. We show an independently measured detector bandwidth curve in Fig.~\ref{fig:PDBW}, as well as a fit to a single-pole low-pass filter (orange line) with a \SI{3}{\decibel} corner frequency at \SI{6.30(14)}{\mega\hertz}. It is obtained by tuning the applied EOM modulation frequency of the probe light, generating an optical beat of constant magnitude. We use the fitted low-pass filter response to remove the detector contribution to the data of Fig.~\ref{fig:PM-SA}(b-c), revealing the intrinsic atomic response.

\bibliography{RydbergSupremacy}% Produces the bibliography via BibTeX.

\end{document}